\documentclass[11pt,a4paper]{article}

\usepackage[left=2.5cm, right=2.5cm, top=2cm, bottom=2cm]{geometry}

\usepackage{authblk}
\usepackage{graphicx} 
\usepackage{natbib}
\usepackage{amsmath,amssymb}
\usepackage{mathtools, amsthm}
\usepackage{tikz}
\usetikzlibrary{arrows.meta, positioning}
\usepackage{tikz-qtree}
\usepackage{enumitem}
\setlist[itemize]{label=\textbullet}
\usepackage{multirow}
\usepackage{comment}
\usepackage{array}
\usepackage[T1]{fontenc}
\usepackage{textcomp}
\usepackage{pgfplots}
\pgfplotsset{compat=1.18}
\usepgfplotslibrary{fillbetween}
\usepackage{iftex}
\usepackage{subcaption}
\usepackage[ruled,vlined,linesnumbered]{algorithm2e}
\usepackage[dvipsnames]{xcolor}
\definecolor{darkblue}{rgb}{0.0, 0.0, 0.55}
\definecolor{chicago-maroon}{RGB}{128,0,0}
\usepackage[bookmarks, colorlinks=true, plainpages = false, citecolor= darkblue, linkcolor = chicago-maroon, anchorcolor = red, urlcolor = darkblue]{hyperref}

\title{Order Dependence in Regression by Composition \\
\large{Discussion on ``Regression by Composition'' by Farewell, Daniel, Stensrud, and Huitfeldt}}
\author[1]{Mei Dong \thanks{may.dong@mail.utoronto.ca}} 
\author[2]{Linbo Wang \thanks{linbo.wang@utoronto.ca}}
\author[3]{Lin Liu \thanks{linliu@sjtu.edu.cn}}
\author[4]{Oliver Dukes \thanks{oliver.dukes@ugent.be}}
\affil[1]{Division of Biostatistics, University of Toronto, Toronto, Canada}
\affil[2]{Department of Statistical Sciences, University of Toronto, Toronto, Canada}
\affil[3]{Institute of Natural Sciences, Shanghai Jiao Tong University, Shanghai, China}
\affil[4]{Department of Mathematics, Computer Science, and Statistics, Ghent University, Ghent, Belgium}
\date{}

\usepackage{setspace}
\begin{document}
 
\maketitle
We congratulate the authors on their inspiring work in developing a new framework in which the conditional distribution is built sequentially, thereby allowing mechanistic structure to be incorporated directly into the model. A key implication, however, is that the framework is inherently order-dependent: changing the order of the flows may lead to different conditional distributions, different parameter interpretations, and different estimation problems.

\begin{figure}
    \centering
    \includegraphics[width=1\linewidth]{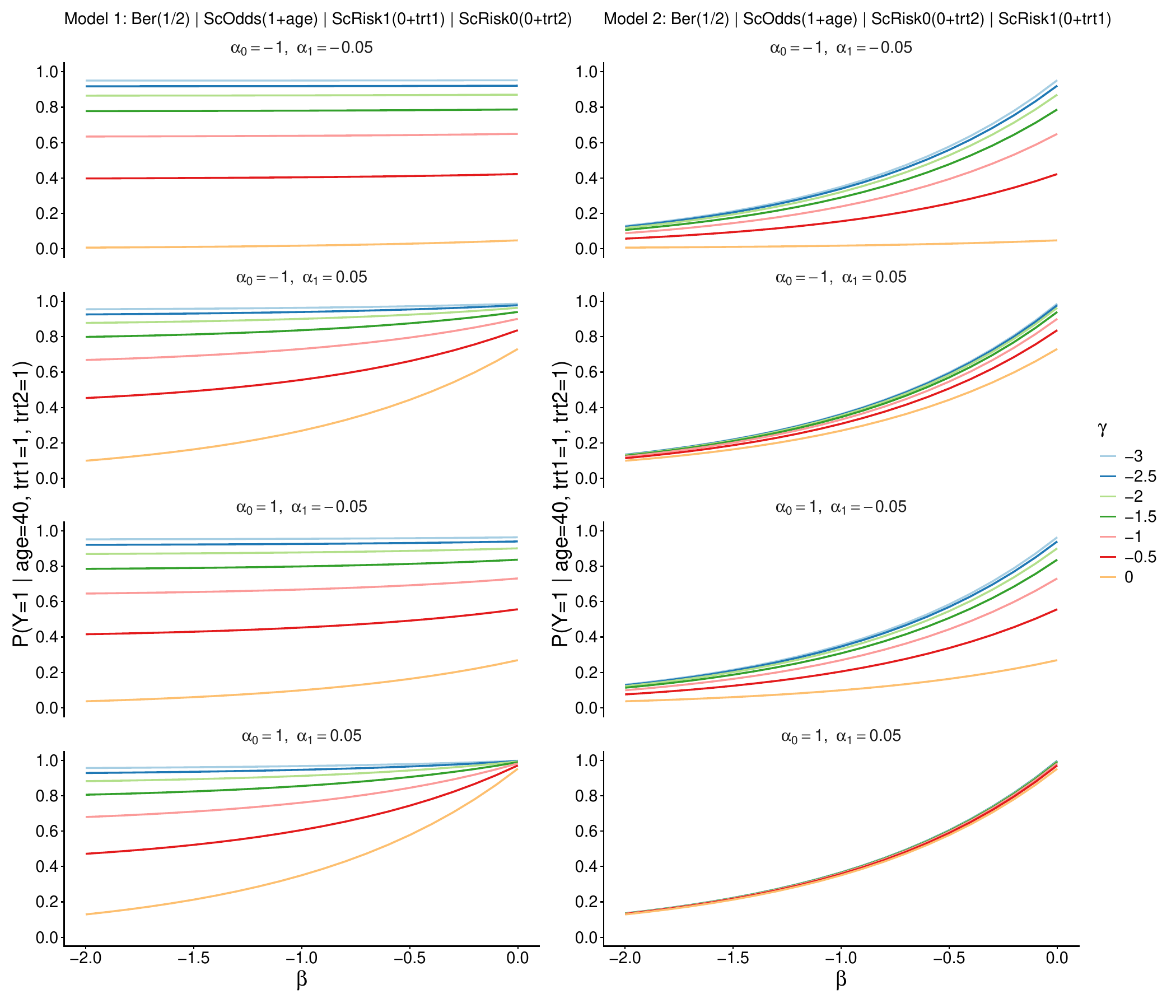}
    \caption{Implied conditional probabilities under Model 1 (left panel) and Model 2 (right panel) across different parameter values, with age fixed at 40 and $\mathrm{trt1}=\mathrm{trt2}=1$. $\alpha_0$ and $\alpha_1$ are the coefficients in the linear predictor $\mathrm{ScOdds}(1+\mathrm{age})$, while $\beta$ and $\gamma$ are the coefficients in the linear predictors $\mathrm{ScRisk1}(0+\mathrm{trt1})$ and $\mathrm{ScRisk0}(0+\mathrm{trt2})$, respectively.}
    \label{fig:impliedprob}
\end{figure}

To illustrate, consider two models with the same components but different orderings:
\begin{itemize}
    \item Model 1: $\mathtt{y=Ber(1/2) \ | \ ScOdds(1+age) \ | \ ScRisk1(0+trt1) \ | \ ScRisk0(0+trt2)}$;
    \item Model 2: $\mathtt{y=Ber(1/2) \ | \ ScOdds(1+age) \ | \ ScRisk0(0+trt2) \ | \ ScRisk1(0+trt1)}$.
\end{itemize}
Suppose $\mathrm{trt1}$ and $\mathrm{trt2}$ are intended to encode a relative risk effect and a survival ratio effect, respectively. Under this framework, the interpretation of $\mathrm{trt1}$ and $\mathrm{trt2}$ depends on the ordering of the flows. The two models imply
\begin{align*}
     &\text{Model 1:} \quad \Pr(Y=1|\text{age}, \text{trt1}, \text{trt2}) = 1-\dfrac{1+\eta_1(\text{age})-\eta_1(\text{age})\eta_2(\text{trt1})}{1+\eta_1(\text{age})}\eta_3(\text{trt2}),\\
    &\text{Model 2:} \quad  \Pr(Y=1|\text{age}, \text{trt1}, \text{trt2}) = \dfrac{1+\eta_1(\text{age})-\eta_3(\text{trt2})}{1+\eta_1(\text{age})}\eta_2(\text{trt1}),
\end{align*}
where $\eta_1(\text{age}) = \exp(\alpha_0 + \alpha_1\text{age})$, $\eta_2(\text{trt1}) = \exp(\beta\,\text{trt1})$, and $\eta_3(\text{trt2}) = \exp(\gamma\,\text{trt2})$. Thus, even with the same building blocks, different orderings induce different parametric models; see Figure \ref{fig:impliedprob} for an illustration.

Order dependence also affects parameter interpretation. In Model 1, although $\exp(\beta)$ is the conditional RR for $\mathrm{trt1}$ under the subcomposition $\mathtt{Ber(1/2) \ | \ ScOdds(1+age)\ |\ ScRisk1(0+trt1)}$, this interpretation is generally not preserved after appending $\mathtt{ScRisk0(0+trt2)}$. Marginalizing over $\mathrm{trt2}$ recovers $\exp(\beta)$ only under specific conditions, derived in the Appendix; in general, it does not. More broadly, later flows may alter the effect measure associated with earlier inputs, and repeated use of the same input across flows may induce composite contrasts rather than standard effect measures. Detailed examples are given in the Appendix.

This also has consequences for estimation. For example, in some orderings, a parameter may retain a familiar interpretation and may admit a doubly robust estimator \citep{richardson2017modeling}. In others, the interpretation may change, and whether a doubly robust estimator exists can likewise depend on the ordering. Hence the ordering of the flows may matter not only for model specification, but also for interpretation and inference.

\bibliographystyle{apalike}
\bibliography{ref.bib}

\newpage

\section*{Appendix}
\subsection*{Examples of order dependence affecting parameter interpretation}
In Model 1, under the subcomposition $\mathtt{ber(1/2) \ | \ ScOdds(1+age)\ |\ ScRisk1(0+trt1)}$,  the conditional RR for $\mathrm{trt1}$ is 
\begin{align*}
    &\mathrm{RR}_{\text{trt1}}(\text{age})=\dfrac{\Pr(Y=1|\text{age}, \text{trt1}=1)}{\Pr(Y=1|\text{age}, \text{trt1}=0)} =\exp(\beta),
\end{align*}
and under the full model, the conditional RR for $\mathrm{trt1}$ is
\begin{align*}
    \mathrm{RR}_{\text{trt1}}(\text{age},\text{trt2})&=\dfrac{\Pr(Y=1|\text{age}, \text{trt1}=1, \text{trt2})}{\Pr(Y=1|\text{age}, \text{trt1}=0, \text{trt2})}\\
    &=\dfrac{1+\eta_1(\text{age})-\eta_3(\text{trt2})+\eta_1(\text{age})\eta_3(\text{trt2})(\exp(\beta) -1)}{1+\eta_1(\text{age})-\eta_3(\text{trt2})}.
\end{align*}

In Model 2, under the full model, the conditional RR for $\mathrm{trt2}$ is
\begin{align*}
    \mathrm{RR}_{\text{trt2}}(\text{age},\text{trt1})&=\dfrac{\Pr(Y=1|\text{age}, \text{trt1}, \text{trt2}=1)}{\Pr(Y=1|\text{age}, \text{trt1}, \text{trt2}=0)} =\exp(\gamma).
\end{align*}

Consider Model 3, in which $\mathrm{trt2}$ appears twice:
\begin{itemize}
    \item Model 3: $\mathtt{y=Ber(1/2) \ | \ ScOdds(1+age) \ | \ ScRisk1(0+trt2) \ | \ ScRisk0(0+trt2)}$.
\end{itemize}
The implied probability is
\begin{align*}
    \Pr(Y=1|\text{age}, \text{trt2}=1) 
    &= 1-\exp{(\gamma)}+\exp{(\beta+\gamma)}\Pr(Y=1|\text{age}, \text{trt2}=0).
\end{align*}
The effect measure of trt2 in Model 3 cannot be naturally interpreted as either a conditional RR or a conditional survival ratio, but rather as a composite treatment contrast. 

\subsection*{Necessary and sufficient conditions under which marginalizing  over $\mathrm{trt2}$ recovers $\exp(\beta)$ in Model 1}

The conditional distribution under Model 1 is
\begin{align*}
\Pr(Y=1\mid \mathrm{age},\mathrm{trt1},\mathrm{trt2})
&=
1-\dfrac{1+\eta_1(\mathrm{age})-\eta_1(\mathrm{age})\eta_2(\mathrm{trt1})}{1+\eta_1(\mathrm{age})}\eta_3(\mathrm{trt2}) \\
&=
1-\eta_3(\mathrm{trt2})
+\eta_3(\mathrm{trt2})\dfrac{\eta_1(\mathrm{age})}{1+\eta_1(\mathrm{age})}e^{\beta\,\mathrm{trt1}}.
\end{align*}
Marginalizing over $\mathrm{trt2}$, we obtain
\begin{align*}
\Pr(Y=1\mid \mathrm{age},\mathrm{trt1})
&=
1-\mathbb{E}\left[\eta_3(\mathrm{trt2})\mid \mathrm{age},\mathrm{trt1}\right]
+\dfrac{\eta_1(\mathrm{age})}{1+\eta_1(\mathrm{age})}e^{\beta\,\mathrm{trt1}}
\mathbb{E}\left[\eta_3(\mathrm{trt2})\mid \mathrm{age},\mathrm{trt1}\right].
\end{align*}
The conditional relative risk for $\mathrm{trt1}$ therefore becomes
\begin{align*}
\dfrac{\Pr(Y=1\mid \mathrm{age},\mathrm{trt1}=1)}{\Pr(Y=1\mid \mathrm{age},\mathrm{trt1}=0)}
&=
\dfrac{
1-\mathbb{E}[\eta_3(\mathrm{trt2})\mid \mathrm{age},\mathrm{trt1}=1]
+\dfrac{\eta_1(\mathrm{age})}{1+\eta_1(\mathrm{age})}e^\beta
\mathbb{E}[\eta_3(\mathrm{trt2})\mid \mathrm{age},\mathrm{trt1}=1]
}{
1-\mathbb{E}[\eta_3(\mathrm{trt2})\mid \mathrm{age},\mathrm{trt1}=0]
+\dfrac{\eta_1(\mathrm{age})}{1+\eta_1(\mathrm{age})}
\mathbb{E}[\eta_3(\mathrm{trt2})\mid \mathrm{age},\mathrm{trt1}=0]
}.
\end{align*}
Thus,
\begin{align*}
\dfrac{\Pr(Y=1\mid \mathrm{age},\mathrm{trt1}=1)}{\Pr(Y=1\mid \mathrm{age},\mathrm{trt1}=0)}=e^\beta
\end{align*}
if and only if
\begin{align*}
&1-\mathbb{E}[\eta_3(\mathrm{trt2})\mid \mathrm{age},\mathrm{trt1}=1]
-e^\beta
+e^\beta \mathbb{E}[\eta_3(\mathrm{trt2})\mid \mathrm{age},\mathrm{trt1}=0] \\
&+\dfrac{\eta_1(\mathrm{age})}{1+\eta_1(\mathrm{age})}e^\beta
\left\{
\mathbb{E}[\eta_3(\mathrm{trt2})\mid \mathrm{age},\mathrm{trt1}=1]
-\mathbb{E}[\eta_3(\mathrm{trt2})\mid \mathrm{age},\mathrm{trt1}=0]
\right\}
=0.
\end{align*}
Since $\mathrm{trt2}$ is binary,
\begin{align*}
\mathbb{E}[\eta_3(\mathrm{trt2})\mid \mathrm{age},\mathrm{trt1}]
&=
\mathbb{E}[e^{\gamma\,\mathrm{trt2}}\mid \mathrm{age},\mathrm{trt1}] \\
&=
1-\Pr(\mathrm{trt2}=1\mid \mathrm{age},\mathrm{trt1})
+e^\gamma \Pr(\mathrm{trt2}=1\mid \mathrm{age},\mathrm{trt1}) \\
&=
1+(e^\gamma-1)\Pr(\mathrm{trt2}=1\mid \mathrm{age},\mathrm{trt1}).
\end{align*}
It then follows that 
\begin{align*}
\dfrac{\Pr(Y=1\mid \mathrm{age},\mathrm{trt1}=1)}{\Pr(Y=1\mid \mathrm{age},\mathrm{trt1}=0)}=e^\beta
\end{align*}
if and only if
\begin{align*}
&(e^\gamma-1)\left\{
e^\beta\left(1-\dfrac{\eta_1(\mathrm{age})}{1+\eta_1(\mathrm{age})}\right)
\Pr(\mathrm{trt2}=1\mid \mathrm{age},\mathrm{trt1}=0)\right. \\
&\qquad\qquad\left.
-\left(1-\dfrac{\eta_1(\mathrm{age})}{1+\eta_1(\mathrm{age})}e^\beta\right)
\Pr(\mathrm{trt2}=1\mid \mathrm{age},\mathrm{trt1}=1)
\right\}
=0.
\end{align*}
Equivalently,
\begin{align*}
(e^\gamma-1)\left[
e^\beta \Pr(\mathrm{trt2}=1\mid \mathrm{age},\mathrm{trt1}=0)
-
\left\{1-\eta_1(\mathrm{age})(e^\beta-1)\right\}
\Pr(\mathrm{trt2}=1\mid \mathrm{age},\mathrm{trt1}=1)
\right]
=0,
\end{align*}
which holds if and only if either $\gamma=0$, so that $\mathrm{ScRisk0}(0+\mathrm{trt2})$ is the identity flow, or the conditional distribution of $\mathrm{trt2}$ satisfies $e^\beta \Pr(\mathrm{trt2}=1\mid \mathrm{age},\mathrm{trt1}=0)
=
\left\{1-\eta_1(\mathrm{age})(e^\beta-1)\right\}
\Pr(\mathrm{trt2}=1\mid \mathrm{age},\mathrm{trt1}=1)$.

\end{document}